\documentclass[reprint,superscriptaddress, amsmath,amssymb, aps, pra, longbibliography]{revtex4-1}

\usepackage{soul}
\usepackage{amsmath}   
\usepackage{amssymb}
\usepackage{mathtools}
\usepackage{graphicx}
\usepackage{dcolumn}
\usepackage{bm}
\usepackage{amsmath}
\usepackage{graphicx}
\usepackage{amsfonts}
\usepackage{subfigure}
\usepackage{graphicx}
\usepackage{array}
\usepackage{float}
\usepackage{color}
\usepackage{multirow}
\usepackage{amssymb}%
\usepackage[colorlinks=true,linkcolor=blue]{hyperref}%
\hypersetup{allcolors=blue}
\usepackage[normalem]{ulem}
\usepackage{xcolor}

\newcommand{\ket}[1]{\ensuremath{\left\vert #1 \right\rangle}}
\usepackage{mathtools}
\DeclarePairedDelimiterX\braket[2]{\langle}{\rangle}{#1 \delimsize\vert #2}

\begin{document}
\title{Effects of inert background gases and photo-illumination on three-color electromagnetically induced transparency of rubidium vapor}

\date{\today }

\author{Alisher Duspayev}
    \email{alisherd@umich.edu}
    \thanks{Present address: Department of Physics and Joint Quantum Institute, University of Maryland, College Park, MD 20742, USA}    
    \affiliation{Department of Physics, University of Michigan, Ann Arbor, MI 48109, USA}
\author{Bineet Dash}
    \affiliation{Department of Physics, University of Michigan, Ann Arbor, MI 48109, USA}       
\author{Georg Raithel}
    \affiliation{Department of Physics, University of Michigan, Ann Arbor, MI 48109, USA} 

\begin{abstract}
Three-color Rydberg electromagnetically induced transparency (EIT) of room-temperature Rb vapor in the presence of inert gases (Ar, Ne, and N$_2$) at 50~mTorr and 500~mTorr is investigated. The observed EIT lines shift  
and develop blue-detuned satellite lines, dependent on inert-gas species and pressure. The separations of the satellite from the main EIT lines are approximately pressure-independent, while their strength increases with inert-gas pressure. The satellite lines are attributed to hyperfine collisions of the intermediate $5D_{3/2}$ state.  Further, analyzing the Stark effect of Rydberg levels, it is found that the inert gases suppress static electric fields in the vapor cells, which we induce by photo-illumination of the cell walls with an auxiliary 453-nm laser. In the work, we utilize Rydberg levels with principal quantum numbers  $n=25$ and $50$ and angular momenta $\ell = 3$ up to 6, excited by the EIT lasers and optional radio-frequency dressing fields.
The work is of interest in the spectroscopic study of mixed-species warm vapors, in sensing applications of Rydberg atoms in vapor cells, and in non-invasive electric-field diagnostics of low-pressure discharge plasma.  
\end{abstract}

\maketitle

\section{Introduction}
\label{sec:intro}

Atomic gases of mixed species are an interesting platform for studies on many-body physics~\cite{Lewenstein2007}, ultracold chemistry~\cite{ulmanis2012}, and precision spectroscopy~\cite{safronovarmp}. They also have a strong connection with various areas of quantum science and technology, with prominent examples including sympathetic cooling of ions~\cite{larson1986, willitsch2008, Hudson2016} and atoms and buffer-gas cooling of molecules~\cite{hutzler2012}. Most of the aforementioned works involve ultra-high-vacuum setups and laser-cooled atomic or molecular samples. Atomic species frequently employed as buffer gases are also used in plasma devices that surround us on a daily basis~\cite{harry2013, eden2013, Nijdam_2022}, such as discharge lamps and waste-processing reactors~\cite{GOMEZ2009614}. With ongoing research and development in these directions, measurement of the electric fields in low-pressure plasma is a relevant challenge~\cite{Goldberg_2022}. Commonly utilized and commercially available field sensors (such as Langmuir probes) may perturb the plasma fields~\cite{Godyak_2011} and have limited spatial resolution. Non-invasive spectroscopic methods add capabilities in spatio-temporal plasma electric-field diagnostics~\cite{alexiou1995, feldbaum2002, park2010, weller2019, viray2020, Mirzaee2020}.

Rydberg-atom-based electromagnetic-field sensing~\cite{9748947, Yuan_2023} could be a suitable candidate for this task. Electromagnetically induced transparency (EIT) in room-temperature alkali vapors~\cite{boller1991, mohapatra2007, kubler2010} as a readout method, combined with the large sensitivity of Rydberg atoms to external fields~\cite{gallagher}, avoids the extensive experimental infrastructure of laser-cooling setups, is compatible with technical plasma environments, and allows traceable electric-field measurements. EIT is conducive to designs for portable self-calibrated field probes~\cite{Sedlacek2012, Holloway2014}. Recent demonstrations of Rydberg-atom-based sensing of stochastic micro-fields within in-vacuum ion sources~\cite{ionsourcepaper}, of photo-induced electric fields in room temperature vapor cells~\cite{ma20, duspayev2023highell}, and of electric fields in photo-ionized alkali vapor~\cite{anderson2017, weller2019} suggest that this method will also be suitable for plasma-field diagnostics in low-pressure inert-gas plasma.

Adding inert gases to alkali-atom vapor cells is often utilized to increase hyperfine and spin coherence times~\cite{vanier1974,balabas2010} and narrow down transition linewidths~\cite{dicke1953, kitching2018}. Most previous experiments on EIT in an inert-gas environment were performed with lowly-excited states~\cite{Novikova:05,sargsyan2010, Cheng_2017, Long:17}. Recently, observations of Rydberg EIT in the presence of inert gases have been reported~\cite{thaicharoen2024, 11cells}. The effects of hyperfine structure and collisional dephasing in three-color Rydberg EIT in a mixture of Rb vapor with low-pressure Ar gas have been studied~\cite{hfspaper2025}. 

Here we investigate three-color Rydberg EIT~\cite{carr2012, Moore2019a} of Rb vapor in the presence of Ar, Ne, and N$_2$ at pressures of 50 and 500~mTorr. 
Furthermore, to explore the capability of such a system for dc electric-field sensing in the inert-gas background, we illuminate the vapor-cell walls to generate photoelectric fields~\cite{ma20, duspayev2023highell}, and investigate the resultant changes in the 
Rydberg-EIT spectra without and with inert gases in the cells.

Our paper is organized as follows. In Sec.~\ref{sec:EIT} we present experimental details and compare three-color Rydberg-EIT in cells without and with inert gas. Using the Stark effect of Rydberg states as a probe, in Sec.~\ref{sec:shield} we then explore and discuss how a low-pressure inert gas alters photo-induced static electric fields in a cell. The paper is concluded in Sec.~\ref{sec:disc}.

\section{Three-photon Rydberg-EIT in mixtures of Rb vapor and inert gases}
\label{sec:EIT}

\subsection{Experimental setup}
\label{subsec:setup}

\begin{figure}[htb]
 \centering
  \includegraphics[width=0.48\textwidth]{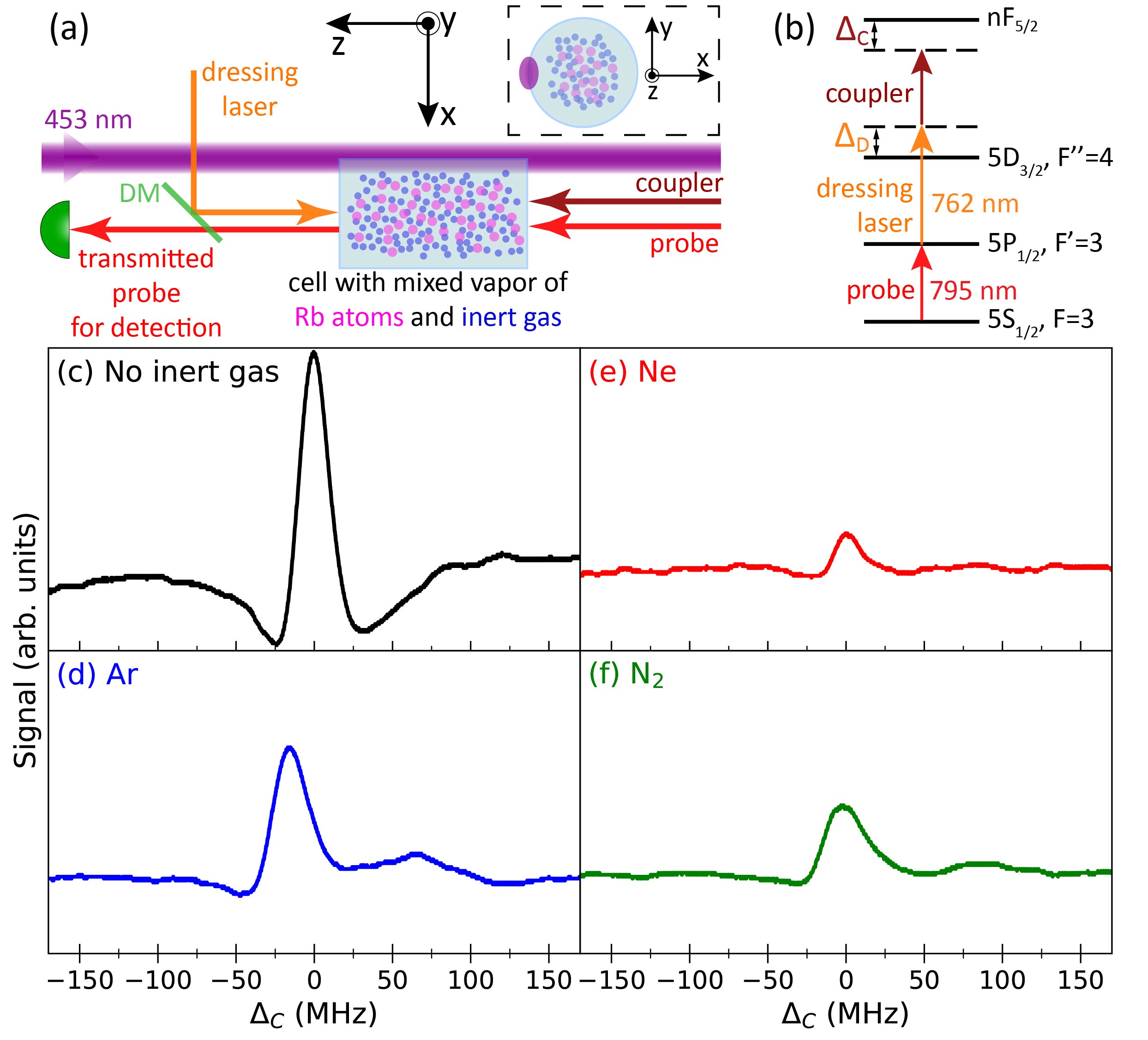}
  \caption{(a) Sketch of the experimental setup, with a cross-sectional view provided in the inset (DM = dichroic mirror). (b) Energy-level diagram of $^{85}$Rb for three-color Rydberg excitation. (c)-(f) Experimental three-color Rydberg EIT signals for cells without inert gas, and with Ar, Ne, and N$_2$, respectively. Vertical scales are identical. In (d)-(f), the nominal pressure of the respective inert gases is 50~mTorr.} 
  \label{fig:setup}
\end{figure}

Our experiments are performed with $^{85}$Rb vapor in room-temperature glass cells. To excite them to Rydberg states, three infra-red lasers (referred to as probe, dressing and coupler) are overlapped within a vapor cell as shown in Fig.~\ref{fig:setup}~(a). The energy-level scheme is shown in Fig.~\ref{fig:setup}~(b). All lasers are cat-eye external-cavity diode lasers (ECDLs). The probe has a wavelength of 795~nm, a power of $\approx$~12~$\mu$W beam, is focused down to a full-width-at-half-maximum (FWHM) of the intensity of $\approx$~80~$\mu$m, and is kept on resonance with the transition $\ket{5S_{1/2}, F = 3} \rightarrow \ket{5P_{1/2}, F' = 3}$ at $\approx$~795~nm. The dressing beam (wavelength 762~nm, power 12~mW, FWHM $\approx$~250~$\mu$m) is frequency-locked at a detuning $\Delta_D \approx$~44~MHz from the transition $\ket{5P_{1/2}, F = 3'} \rightarrow \ket{5D_{3/2}, F'' = 4}$. Two coupler lasers with wavelengths of 1280~nm and 1255~nm are employed to excite Rydberg levels with principal quantum numbers $n = 25$ and $n = 50$, respectively. The coupler lasers are passed through a shared optical amplifier, one at a time, avoiding coupler-laser realignment when switching $n$. The amplifier output is focused into the vapor cells with a FWHM spot size of $>$~250~$\mu$m. We achieve $\approx$~7 and 10~mW of power for the 1280-nm and 1255-nm couplers, respectively, before the vapor cells. A small sample of coupler-laser light is sent through a reference Fabry-P\'erot (FP) etalon of 374~MHz free spectral range (FSR). FP transmission functions are recorded simultaneously with the Rydberg-EIT spectra, as the coupler detuning $\Delta_C$ is scanned across the Rydberg states of interest. The peaks in the FP functions are used to calibrate and linearize the coupler frequency scans. For improved accuracy and linearity, we RF-modulate the coupler beam sample using an electro-optic modulator to increase the number and density of frequency calibration peaks in the FP functions. Probe and coupler beams are co-aligned in the $+z$-direction, whereas the dressing beam is counter-aligned in the $-z$ direction [see Fig.~\ref{fig:setup}~(a)]. All lasers are linearly polarized parallel to $x$-axis. More details are provided in~\cite{duspayev2023highell, hfspaper2025}.

The three-color Rydberg-EIT signal is revealed in the acquired transmission spectrum of the probe laser [see Fig.~\ref{fig:setup}~(a)]. To improve signal-to-noise ratio (SNR), we implement lock-in detection by amplitude-modulating the coupler beam using an acousto-optic modulator (AOM). The AOM pulse-modulates the coupler with a 3~kHz square pulse at a 50\% duty cycle. Each spectrum presented is an average over typically 400 scans of the lock-in amplifier output versus $\Delta_C$, where the average is taken on an oscilloscope. An example of a three-color Rydberg-EIT signal in a cell without inert gas is shown in Fig.~\ref{fig:setup}~(c) for the Rydberg state $25F_{5/2}$. The linewidth of~$\approx$~22~MHz achieved in Fig.~\ref{fig:setup}~(c) results from a combination of 
power broadening, laser linewidths, frequency jitter of probe and dressing lasers relative to their respective lock points, and frequency jitter of the coupler laser during the scans. 

To create static electric fields in the cells in a controlled fashion, we employ the photoelectric effect on the cell walls~\cite{hankin2014, ma20, duspayev2023highell, jau2020, patrick2025} using an additional 453-nm laser beam of $\approx$~1-cm diameter. The 453-nm beam propagates in the $xz$-plane and, due to spatial constraints, forms an angle of $\approx$~20$^\circ$ with respect to the cell axis ($z$-axis). The 453-nm beam strikes a well-defined region on the cell wall, as shown in Fig.~\ref{fig:setup}~(a). The generated photoelectric field points along the $x$-direction~\cite{ma20}.

\begin{figure*}[t!]
 \centering
  \includegraphics[width=0.75\textwidth]{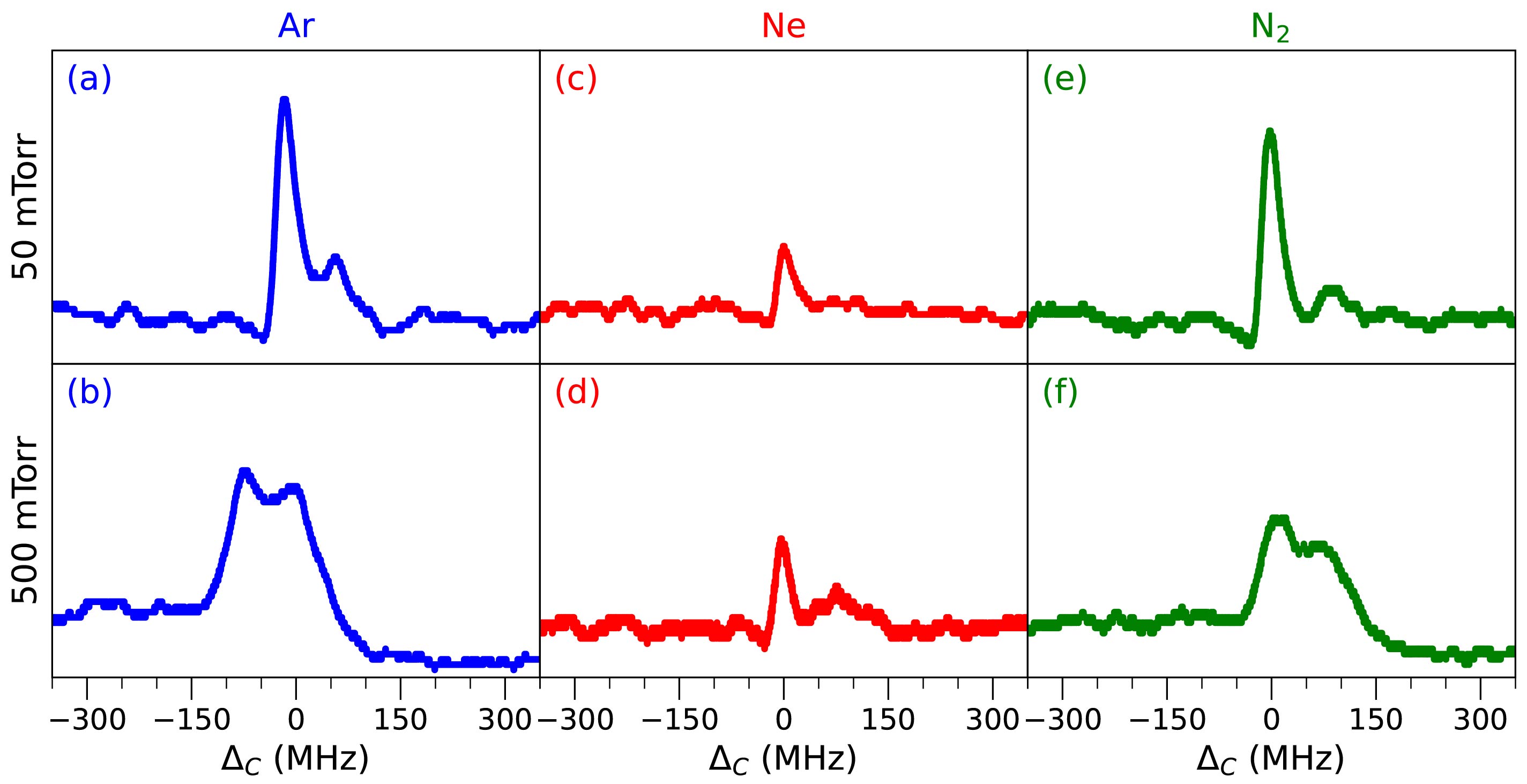}
  \caption{Three-color Rydberg-EIT signals at 50~mTorr (top row) and 500 mTorr (bottom row) of (a)-(b) Ar, (c)-(d) Ne, and (e)-(f) N$_2$ inert gas in the cells. The $\Delta_C$- and $y$-scales are the same in all plots.} 
  \label{fig:pressure}
\end{figure*}

\subsection{Three-color Rydberg-EIT in the presence of inert gases}
\label{subsec:res1}

After obtaining a reference EIT signal with the inert-gas-free vapor cell [Fig.~\ref{fig:setup}~(c)], in Figs.~\ref{fig:setup}~(d)-(f) we exchange the cell with one out of three cells containing Ar, Ne or N$_2$ at 50~mTorr. 
The utilized Rydberg state is $25F_{5/2}$, the same as in Fig.~\ref{fig:setup}~(c). The $y$-axes in Figs.~\ref{fig:setup}~(c)-(f) are identical to allow for a direct comparison of signal strengths. After each cell exchange, we fine-adjust the coupler and the dressing beams to re-optimize the EIT signal after cell replacement. Typically, the fine-adjustment yields only minor improvements, as expected from the optical quality of the cells. We observe an EIT signal in all three cases, as shown in Figs.~\ref{fig:setup}~(d)-(f).  The main observations are as follows.

The EIT signals from the inert-gas cells are notably smaller than those of the reference cell in Fig.~\ref{fig:setup}~(c). For Ar, Ne, and N$_2$, the EIT peak heights are reduced by factors of $\approx$~2, $\approx$~7 and $\approx$~4, respectively. However, due to cell availability, the cells with inert gases are~$\approx$~4~cm long, while the inert-gas-free cell is~$\approx$~7.5~cm long. Taking into account cell length, we estimate that 50~mTorr of Ar gas does not significantly reduce the EIT signal strength, whereas 50~mTorr of Ne and N$_2$ reduce the EIT signal by factors between about 3 and 2, respectively. 

The measured EIT line shifts in Fig.~\ref{fig:setup}~(d)-(f) are $\approx$~-15~MHz, $\approx$~+1~MHz, and $\approx$~-1~MHz for Ar, Ne, and N$_2$, respectively; only in the case of Ar is the shift significantly larger than the statistical uncertainty of $\sim 2$~MHz. These results are consistent with our previous study using two-color Rydberg EIT~(see~\cite{11cells} and references therein). 

The FWHM linewidths of the main EIT peaks for Ar and N$_2$ [Figs.~\ref{fig:setup}~(d) and (f)] are determined to be~$\approx$~28~MHz and~31~MHz, respectively, indicating a slight EIT line broadening caused by the presence of Ar or N$_2$. For the case of Ne [Fig.~\ref{fig:setup}~(e)], the observed FWHM linewidth is~$\approx$~20~MHz, which is same as that without inert gas [Fig.~\ref{fig:setup}~(c)] within our uncertainty of $\approx$~2~MHz. Although a comparison of three-color EIT linewidths with those from our previous two-color study~\cite{11cells} is not feasible over a wide pressure range due to the rise of the satellite lines at higher pressures (see Sec.~\ref{subsec:res2}), in Figs.~\ref{fig:setup}~(d) and (f) we observe trends similar to those previously seen in two-color EIT: at same pressure, Ne causes the least line broadening, while Ar and N$_2$ cause similar and significantly larger broadening.

In the cases of Ar and N$_2$ in Fig.~\ref{fig:setup}, we notice weak, inert-gas-induced satellite lines at~$\approx +60$~MHz and~$\approx +~85$~MHz relative to the main EIT peaks, respectively. In the case of Ne [Fig.~\ref{fig:setup}~(e)], which shows the weakest EIT signal overall, the satellite line is below the noise level. 

\subsection{Effect of inert-gas pressure}
\label{subsec:res2}

Three-color Rydberg EIT signals are also observed in vapor cells with an inert-gas pressure of 500~mTorr. The results are shown in the bottom row in Fig.~\ref{fig:pressure} together with the 50-mTorr data (top row) for comparison. Attempts to obtain three-color Rydberg EIT signals at 5~Torr did not lead to any conclusive results. 

At 500-mTorr of Ar, the main EIT peak and the inert-gas-induced blue-shifted satellite feature morph into a pronounced double-peak structure of near-symmetric peaks of equal amplitude [compare Figs.~\ref{fig:pressure}~(a) and~(b)]. Moreover, an additional overall shift of $\approx-60$~MHz occurs, {\sl{i.e.}}  the main EIT peak (left peak) shifts in $\Delta_C$ by $\approx-75$~MHz from its inert-gas-free position. 

In the case of Ne [Figs.~\ref{fig:pressure}~(c) and~(d)], the main EIT line shifts by about $\approx$~-1~MHz and $\approx$~-2~MHz
at 50~mTorr and 500~mTorr, respectively [Figs.~\ref{fig:pressure}~(c) and~(d)], which are both insignificant. For 500~mTorr of Ne, a comparatively weak inert-gas-induced satellite feature is observed at~$\Delta_C~\approx$~+75~MHz. Apparently, for 50~mTorr of Ne and at our SNR, the satellite is too weak to be observed. 

Finally, in the case of N$_2$, the spectrum at 500~mTorr [Fig.~\ref{fig:pressure}~(f)] exhibits two peaks of near-identical height, similar to Ar, but with a marginally wider satellite (right peak). The stronger, left peak shifts to the blue by a significant amount of $\approx$~10~MHz, whereas the satellite appears to move slightly to the red, causing the two peaks at 500~mTorr [Fig.~\ref{fig:pressure}~(f)] to appear slightly closer to each other than at 50~mTorr [Fig.~\ref{fig:pressure}~(e)]. Due to line pulling, which moves spectral lines closer to each other once they overlap significantly, and our $\sim 2$~MHz uncertainty in the peak positions, we do not believe that this reduction in peak separation is significant.

The signal areas under the main EIT components (left peaks) do not exhibit a significant pressure-dependent change. In the case of Ar and N$_2$, the main EIT peaks broaden by about a factor of two between 50~mTorr and 500~mTorr, while for Ne, the main peak's width and height do not substantially change. These trends are consistent with~\cite{11cells}, where Ne showed a factor of 2 to 3 less line broadening than Ar and N$_2$, which were both similar. Also, because at low pressure the pressure-independent intrinsic EIT linewidth is dominant in all cases, in none of the cases do we see a factor of ten linewidth increase between 50~mTorr and 500~mTorr. 

In Fig.~\ref{fig:pressure}, the areas under the inert-gas-induced satellites increase by about an order of magnitude between 50~mTorr and 500~mTorr of Ar or N$_2$. In recent work~\cite{hfspaper2025}, we explained the Ar-induced satellite line by hyperfine collisions between Rb atoms in the $5D_{3/2}$ state and Ar atoms using a 10-level simulation of the Lindblad equation. The present work generalizes observations of the satellite line to other inert gases and to higher pressure, where the satellite rivals the main EIT line in strength. We have performed several checks to further strengthen our explanation of the origin of the satellite lines. First, varying the frequency of the 795-nm laser [see Fig.~\ref{fig:setup}~(b)] causes the whole spectrum, including the satellite, to shift together. This indicates that the $5P_{1/2}$ state and its hyperfine structure do not cause the satellite. Next, changing the principal quantum number of the Rydberg state, $n$, does not cause the satellite to shift with respect to the main Rydberg EIT line. This indicates that the satellite is not due to Rydberg-atom interactions. Finally, when varying the dressing-beam detuning, $\Delta_D$, the satellite does not significantly shift, whereas the main Rydberg-EIT line shifts linearly by $-\Delta_D$. Here, we have varied $\Delta_D$ over a range of 30~MHz to ascertain this dependence.

Based on the results for three-color EIT signal strength presented in this Section, linewidth and line shape, Ar emerges as a preferred choice for future Rydberg-EIT-based spectroscopic electric-field diagnostics. Specifically, low-pressure Ar plasma may be a good candidate for concurrent plasma operation and Rydberg-EIT electric-field measurement with Rb vapor immersed into the Ar.

\section{Three-color Rydberg EIT with dc electric fields}
\label{sec:shield}  

To probe dc electric fields in vapor cells filled with mixtures of inert gas and Rb atoms, we perform two sets of experiments described in this Section. We induce the static fields using photo-illumination of the vapor cell wall using a 453-nm laser, as described in Sec.~\ref{subsec:setup} and shown in Fig.~\ref{fig:setup}~(a). The 453-nm laser power is $\gtrsim 100$~mW, which yields electric fields up to $\sim 50$~V/m~\cite{ma20, duspayev2023highell}. 

\begin{figure}[htb]
 \centering
  \includegraphics[width=0.44\textwidth]{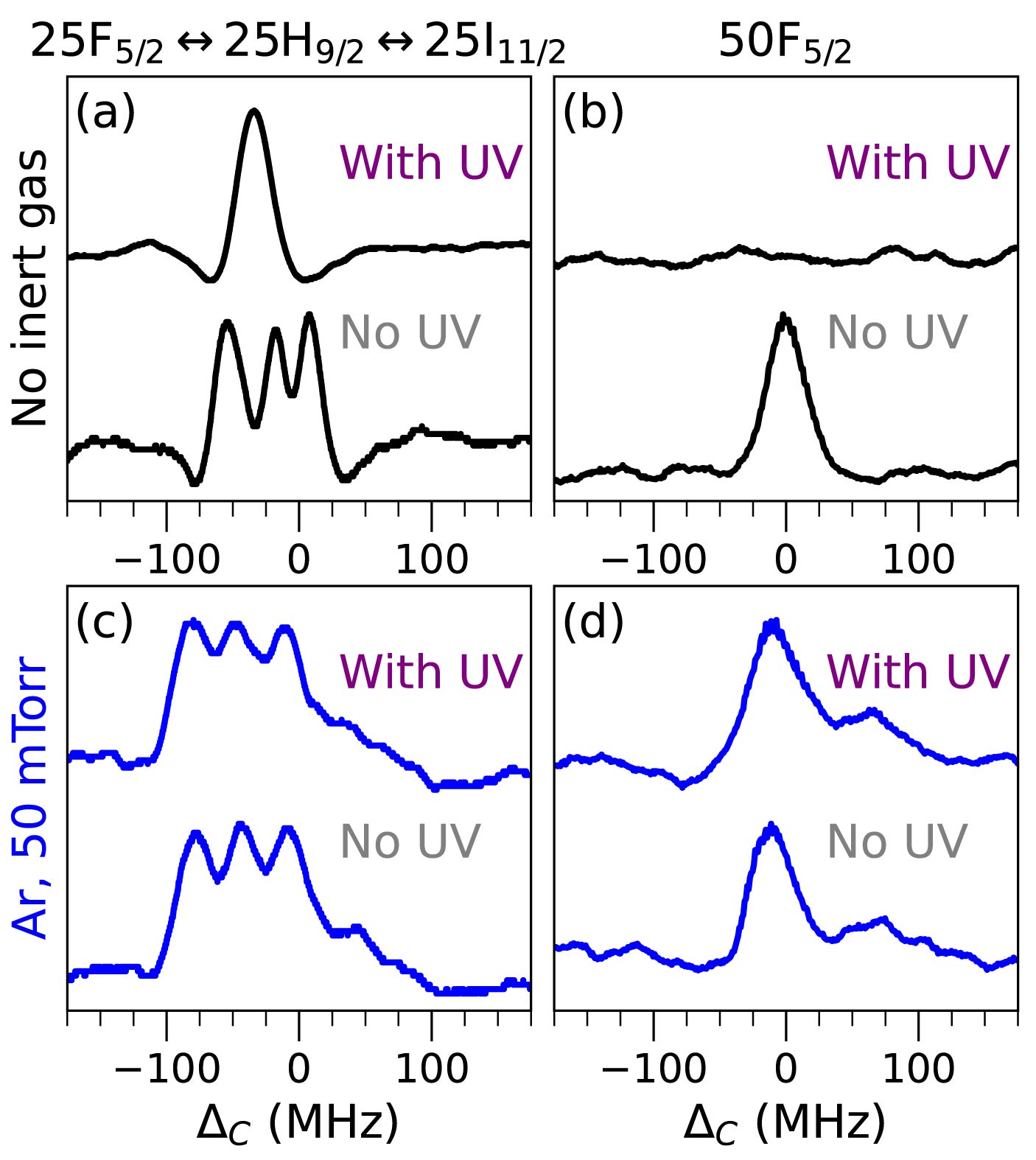}
  \caption{Rydberg EIT with the RF-coupled triplet of states, $25F_{5/2}$, $25H_{9/2}$, and $25I_{11/2}$ (left column) and with the $50F_{5/2}$ state (right column), without and with the 453-nm light shone onto the cell walls. Top row - inert-gas-free cell, bottom row - cell with 50~mTorr of Ar. EIT peaks obtained from the inert-gas-free cell with the 453-nm light and the RF dressing fields turned off were used to mark the location $\Delta_C = 0$.} 
  \label{fig:shield}
\end{figure}

\subsection{$25F_{5/2}\leftrightarrow25H_{9/2}\leftrightarrow25I_{11/2}$ transition}
\label{subsec:lown}  

In our first approach to demonstrate photo-induced dc electric fields using three-color Rydberg-EIT, we utilize additional radio-frequency (RF) dressing fields to couple the high-$\ell$ levels $25H_{9/2}$ and $25I_{11/2}$, which exhibit large electric polarizabilities~\cite{brown2023, duspayev2023highell, prajapati2023highell, allinson2024}. One RF dressing field resonantly drives the two-photon $25F_{5/2}\leftrightarrow25H_{9/2}$ transition at $\approx 2~\times$~3.16~GHz, while  another resonantly drives the one-photon $25H_{9/2}\leftrightarrow25I_{11/2}$ transition at $\approx$~345~MHz. The scheme produces three coherently mixed RF-dressed Rydberg levels that are optically accessible and have large electric polarizabilities due to their high-$\ell$ admixtures. Details of the RF setup are described in~\cite{duspayev2023highell}.

In Fig.~\ref{fig:shield}~(a) we show the three-color EIT spectra of the RF-coupled triplet $25F_{5/2}\leftrightarrow$  $25H_{9/2}\leftrightarrow25I_{11/2}$ in the cell with no inert gas, with both the photo-illumination laser at 453~nm off and on. When the 453~nm laser is turned on, the dressed-state triplet spectrum reduces dramatically to a single peak. This is because the photo-induced dc electric field Stark-shifts the $25H$- and $25I$-levels by $\gtrsim - 100$~MHz, or by significantly more than the estimated RF Rabi frequencies. The $25H$- and $25I$-levels then do not couple any more to $25F$ via the RF dressing fields, leaving only one EIT peak (as observed). 

Quantitatively, the scalar polarizabilities are 0.0100~MHz/(V/m)$^2$, 0.106~MHz/(V/m)$^2$, and 0.248~MHz/(V/m)$^2$ for 25$F$, 25$H$ and 25$I$, respectively. Due to the tensor contributions, the polarizabilities for the respective magnetic sub-levels vary by factors $\lesssim 2.3$. Noting that the 25$H, m_\ell=5$ sub-level has the smallest polarizability among all  $25H$ and $25I$ sub-levels, with $\alpha=0.0626$~MHz/(V/m)$^2$, and that the dc Stark shift is $- \alpha E_{DC}^2/2 $ in an electric field of $E_{DC}$, we conclude that $E_{DC} \gtrsim 50$~V/m. Further, with the 453-nm light on and the RF fields off the $25F$-manifold Stark-shifts from $\Delta_C=0$ to $\Delta_C~\approx-12$~MHz, which, using the scalar polarizability listed above, results in $E_{DC} \sim 50$~V/m. These estimates are in line with our earlier studies~\cite{ma20, duspayev2023highell}.

Analogous $n=25$ Rydberg-EIT spectra using the cell with 50~mTorr of Ar are shown in Fig.~\ref{fig:shield}~(c). There, we increase the power of the RF source driving the $25H_{9/2}\leftrightarrow25I_{11/2}$ transition by 4~dB and shift the frequency of the source driving the $25F_{5/2}\leftrightarrow25H_{9/2}$ transition by 5~MHz. It was found empirically that these minor modifications improve the SNR and enhance the symmetry of the triplet spectra. The satellite features on the blue side are similar to the ones in Fig.~\ref{fig:pressure} and are interpreted by hyperfine collisions of the $5D_{3/2}$ state, which has no significant electric-field sensitivity. Interestingly, the three-peak EIT signal remains almost unchanged when the 453-nm laser is applied. A notable change in Fig.~\ref{fig:shield}~(c) relative to the no-UV case in Fig.~\ref{fig:shield}~(a) is a moderate broadening that causes the valleys between the three EIT peaks to fill in more. The broadening includes collision-induced broadening (see Figs.~1 and~2) from the Ar gas. Comparing both signals in Fig.~\ref{fig:shield}~(c), we see a hint of additional broadening caused by the 453-nm light. Results of experiments obtained with Rb cells containing 50-mTorr of Ne or N$_2$ are similar in that the inert gases remove the effect of the 453-nm light on the triplet EIT signals (not shown).

\subsection{$50F_{5/2}$ transition}
\label{subsec:highn} 
To affirm the results from Sec.~\ref{subsec:lown}, we perform an electric-field test using the $50F_{5/2}$ state, without extra RF dressing fields. We utilize a different coupler laser tuned to 1255~nm (see Sec.~\ref{subsec:setup}). The EIT spectra obtained using the inert-gas-free and the 50-mTorr-of-Ar cells are shown in Figs.~\ref{fig:shield}~(b) and (d), respectively. 
In the case of the inert-gas-free cell, the EIT peak is completely gone when the 453-nm laser is applied. Noting that the smallest polarizability of $50F_{5/2}$ is $0.805~$MHz/(V/m)$^2$, for $m_\ell=3$, and that the Stark shifts must exceed $\sim 100$~MHz for the EIT line to vanish, the photo-induced electric field in the inert-gas-free cell is estimated at $E_{DC} \gtrsim 20$~V/m, which is consistent with the results from Sec.~\ref{subsec:lown}

In Fig.~\ref{fig:shield}~(d) we observe, again, an inert-gas-induced satellite on the blue side, similar to Fig.~\ref{fig:pressure}. Moreover, with the addition of 50~mTorr of Ar the EIT line remains when the 453-nm laser is turned on, showing that the photo-induced electric field caused by the 453-nm laser disappears. The $50F_{5/2}$-line may be slightly Stark-shifted by an amount $\lesssim 5$~MHz, which sets an upper limit of about $3$~V/m to the remnant photo-induced electric field in the EIT excitation region. Combining this result with the estimate of $50~$V/m from Sec.~\ref{subsec:lown} for the photo-induced electric field in inert-gas-free cells, we conclude that 50~mTorr of Ar reduces the photo-induced electric field by a factor $\gtrsim 16$.

The results from this Section show that an inert-gas filling of 50-mTorr inhibits photo-induced electric fields in glass cells made from Pyrex (our case). One may argue that 50-mTorr of Ar (an inert gas) should not affect the photoelectric properties of the glass itself. Under that assumption, one is led to believe that the Ar gas modifies the Rydberg-gas and free-charge diffusion properties in a way that photoelectric fields originating from the glass surfaces are inhibited via a volume charging effect. 

\section{Conclusion and Outlook}
\label{sec:disc}

We have presented an investigation of three-color Rydberg EIT in the presence of three different inert gases at 50 and 500~mTorr. The EIT lines shift and broaden in response to the inert-gas pressure, in agreement with earlier two-color studies, and develop blue-shifted, pressure-dependent satellite features that are attributed to $5D_{3/2}$ hyperfine collisions. Further, we have investigated Stark shifts of three-color Rydberg-EIT lines, induced by the photoelectric effect on the cell walls. The inert gases were found to largely eliminate the photo-induced Stark shifts. 

A natural extension of the presented work includes electric-field sensing in discharge plasmas, enabled by three-color Rydberg-EIT of Rb sensor atoms immersed in a low-pressure inert gas that serves as the plasma medium. Based on the present work, several tens of mTorr of Ar are expected to be a good choice. Applications of spectroscopic electric-field sensing to diagnose artificial and natural plasmas~\cite{moore1984, Goldberg_2015, dogariu2017} may ensue.

A detailed study of the observed inert-gas-induced spectral features in Fig.~\ref{fig:pressure} could be a topic of future investigations~\cite{Zameroski_2014}, including alkali-inert-gas collisions in other EIT media ({\sl{e.g.,}} Cs), other intermediate states with different hyperfine structure ({\sl{e.g.,}} $4D_{J}$ in~\cite{lee2015, lim2022, Duspayev_2023}), and other inert gases. 

Further study will be required to verify and to elucidate the observed suppression of photo-induced electric fields in cells by low-pressure ($\lesssim 50$~mTorr) inert-gas cell fillings. Such experiments could utilize alternative methods of generating static electric fields in vapor cells, such as embedded electrodes~\cite{barredo2013, Grimmel_2015, andersonapl2018, ma2022}. A reduction of perturbing dc fields may be beneficial in Rydberg-atom-based sensors~\cite{Adams_2020, Anderson.RydergCommsSensing.2020,fancher2021, 9748947, Yuan_2023}.

\section*{Acknowledgments}
\label{sec:acknowledgments}
We would like to thank Ryan Cardman for contributing to the construction of the experimental setup, and Nithiwadee Thaicharoen and Eric Paradis for useful discussions. This work was supported by the U.S. Department of Energy, Office of Science, Office of Fusion Energy Sciences under award number DE-SC0023090 and NSF grants Nos. PHY-2110049 and PHY-2412535. A.D. acknowledges support from the Rackham Predoctoral Fellowship at the University of Michigan.

\bibliography{bibliography.bib}

\end{document}